# Experimental Performance Evaluation of Cloud-Based Analytics-as-a-Service

Francesco Pace*, Marco Milanesio*, Daniele Venzano*, Damiano Carra† and Pietro Michiardi*
*Data Science Department, Eurecom, Biot Sophia-Antipolis, France
†Computer Science Department, University of Verona, Verona, Italy
Email: *{name.surname}@eurecom.fr, †damiano.carra@univr.it

*Abstract*—An increasing number of Analytics-as-a-Service (AaaS) solutions has recently seen the light, in the landscape of cloud-based services. These services allow flexible composition of compute and storage components, that create powerful data ingestion and processing pipelines. This work is a first attempt at an experimental evaluation of analytic application performance executed using a wide range of storage service configurations. We present an intuitive notion of data locality, that we use as a proxy to rank different service compositions in terms of expected performance. Through an empirical analysis, we dissect the performance achieved by analytic workloads and unveil problems due to the impedance mismatch that arise in some configurations. Our work paves the way to a better understanding of modern cloud-based analytic services and their performance, both for its end-users and their providers.

## I. Introduction

Large-scale computing frameworks have received a lot of attention recently, as today they constitute essential tools that industries exploit to extract value from their data assets. Thanks to virtualization, compute and storage clusters are more flexible, they can be easily provisioned in different sizes, and destroyed when not needed [1]. Increasingly, such storage and processing systems are exposed to users as *services*, deployed on either public or private cloud computing environments, rather than on bare-metal machines in private clusters. Indeed, many companies offer Analytics-as-a-Service (AaaS) clusters to run a variety of applications: Amazon Web Services (AWS) with Elastic MapReduce [1], DataBricks Cloud[1] [2], Cloudera Cloud[1] [3] and Google Cloud Hadoop [4] are noteworthy examples.

In cloud computing environments, the architecture of analytics clusters is the result of the composition of several services, consisting of three (logically separated) layers: the *Compute layer* refers to all cluster nodes that run the data processing application (e.g., a Spark application); the *Data layer* refers to any combination of storage services (e.g., HDFS [5] or Swift [6]); and the *Storage layer* that physically stores the data, including ephemeral disks, object and elastic block stores.

Currently, users of AaaS have abundant information about pricing, on the one hand, and about the durability of resources, on the other hand. Although a tedious and complex exercise, it is possible to reason about cost-based service dimensioning, and to select appropriate storage services depending on data availability and durability objectives. As a consequence, it is today possible to build data ingestion, storage and processing pipelines, by composing – in various combinations – the three layers defined above.

Surprisingly, as of today, very little work has been done to shed light on the intricate relation that exists between the performance of analytics applications running on such cloud-based services, and their composition. As such, the endeavor of this work is to study the impact of different configurations of Compute, Data and Storage layers on the performance of a data analytics framework, with focus on application runtime. To this aim, we take an experimental approach, and propose a measurement methodology and campaign, whose objective is to analyze the performance corresponding to an intuitive notion of distance between where computation happens and data reside. In doing so, we define an extensive set of application workloads that challenge the systems under study in different ways. Ultimately, our goal is to overcome the limitations of prior works that only provide a boolean vision of data locality: our results indicate that – in general – the intuitive distance metric we present in this work is a good proxy to reason about performance ranking. However, impedance mismatch between different services and application workloads must be taken into account to formulate plausible explanations for outliers in terms of performance.

In summary, the contributions of this work are the following:

- We perform an extensive measurement campaign on a private cloud computing environment, completely under our control. Our study involves the combination of several analytics services. For each deployment scenario, we report and explain the measured performance of a variety of application workloads, including read/write intensive, business intelligence and machine learning applications.
- We present an intuitive notion of data locality that can be used as a proxy to rank different service compositions, in terms of expected performance. We critically examine the validity of our intuition as a function of application workloads, and identify and explain outliers.
- We present experimental evidence of the impedance mismatch between large-scale computing framework and two important storage layers – object stores and elastic block stores – and deduce mechanism to mitigate negative effects on performance.

The rest of the paper is organized as follows: in Section II

---

[1] Solution hosted on Amazon Web Services (AWS)

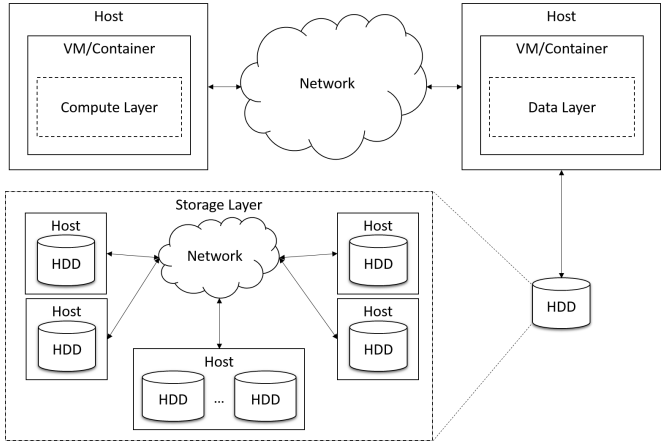

Fig. 1. Logical components in an analytics cluster.

we introduce the terminology and we detail the domain of our work; in Section III we explain our research question and in Section IV we describe our approach. The experimental results and the discussion are presented in Section V. Finally, Section VI summarizes our contributions and presents some future research directions.

## II. BACKGROUND AND RELATED WORKS

### A. Analytic services components

Analytics services consist of three main logical components, as illustrated in Fig. 1:

1) The Compute layer runs parallel processing frameworks that execute analytics applications (or jobs), e.g., Apache Spark [7]. An example of compute layer service is Amazon Elastic MapReduce.
2) The Data layer exposes logical storage services from the underlying physical storage system to the Compute layer. Examples of data layers include Hadoop HDFS [5] and Amazon S3 [8].
3) The Storage layer handles read/write operations and has direct access to the data. Disks exposed to the data layer can be a series of individual, ephemeral devices as well as more complex distributed file systems.

The distinction between Data and Storage layers is a consequence of virtualization. Indeed, a Virtual Machine (VM) hosting the Data layer (e.g., the HDFS' DataNode) might use a virtual disk which resides on a different host. For example, many public cloud providers expose virtual disks provisioned by an underlying distributed file system to improve, among others, VMs' migration time. In general, the Compute layer requires a compatible Data layer to access the Storage layer.

### B. Related Works

The performance implications of data locality have been investigated in several studies. We can identify two major trends: one (e.g., [9]–[11]) arguing that data locality is not relevant, while the other (e.g., [12]–[17]) highlighting the opposite. All these works, albeit valid, base their conclusion on limited information and define data locality as a boolean feature (present or not). We move on from this dichotomy by investigating applications' performance in a variety of service compositions, leading to various degrees of data locality.

Considering the methodology, we can divide the recent research efforts to understand the impacts of data locality into three main categories: *(a)* analysis on limited/public configurations, *(b)* analysis on limited workloads and *(c)* theoretical or trace-driven analysis.

*a) Limited/Public Configurations:* Examples of works that use limited/public deployments can be found in [9]–[13]. For example, Ousterhout et al. in [9] use an ideal scenario in terms of data locality (Compute and Data layer on the same VM), with limited knowledge of the underlying Storage layer. With the help of an analysis performed on network, disk block time and percentages of resource utilization, such work state that the runtime of analytics applications is generally CPU-bound rather than I/O intensive; thus, data locality may be considered irrelevant. We show that this is not always true.

*b) Limited Workloads:* The studies presented in [12], [16], and [18] use a limited set of workloads to investigate data locality. For example, Xie et al. in [12] use two workloads: a WordCount and a Grep-like applications to demonstrate that data placement plays an important role in analytics applications. While this consideration is valid, with our approach, we recognize the importance of workload heterogeneity in studying system performance.

*c) Other limitations:* Some authors base their works on theoretical or trace-driven analysis [10], [13]–[15]. The work from Ananthanarayanan et al. in [10] is based on Facebook's traces. The authors state that, since network technology evolves quickly, data locality is an aspect that will soon be neglected; they also use micro benchmarks to study a single aspect of an analytics cluster. Instead, we show that in some cases the network might indeed not be a bottleneck, while in others it may contribute to harm application performance. In addition, using a micro-benchmark approach alone to measure I/O performance, can lead to inaccurate results, since analytics frameworks like Spark or Hadoop are more complex than a set of read/write operations.

Works like [19]–[21] do not fall directly into one of these categories, as they model different aspects of a MapReduce application. Nonetheless, they leave data locality as an abstract concept and they always consider configurations when Compute, Data and Storage are on the same host. For example, Yang et al. in [19] model the relationship between number of mapper and reducers, while Lin et al. in [20] model an entire analytics application that uses Hadoop and an analytics framework. Zhang et al. in [21] create a model to improve the data locality when Compute, Data and Storage layers are on the same machine. We go further and study data locality when there is a clear separation of Compute and Data layer.

## III. PROBLEM STATEMENT

We empirically evaluate the performance of analytics applications composed using a variety of Compute, Data and

Storage layer configurations. Our goal is to understand how application run time varies across configurations, for a wide range of application workloads.

Performance modeling of complex, distributed systems is a daunting task: application runtime is affected by several factors, including *data locality* (which is the foundation of parallel processing frameworks such as Hadoop and Spark), impedance mismatch between the various services involved in analytic applications, interference between competing tenants, application workloads, and many more.

As such, we take an experimental approach, and analyze application performance through the lenses of the data locality principle, which we revisit to accommodate the breath of storage configurations currently available in most public and private clouds. In our study, we emphasize problems that arise as a consequence of service composition, and suggests ways to mitigate them.

## IV. METHODOLOGY

We use a private cloud computing platform to overcome the limitations of experiments performed on public cloud infrastructures and we run 4 different types of analytics applications: read intensive, write intensive, business intelligence and machine learning. In this section we: *(i)* provide the specifics of our platform, *(ii)* illustrate the different placements of Compute, Data and Storage layers that we use in our experiments, *(iii)* introduce an intuitive notion of distance between where computation happens and data reside, that we call the *Compute-to-Data path*, *(iv)* present our workloads and *(v)* explain the metrics used.

### A. Experimental Platform

Our platform is composed by 25 server-grade machines equipped with: *(i)* two sockets with an Intel Xeon at 2.40GHz, with hyper threading enabled (32 Cores), *(ii)* 128 GB RAM and *(iii)* ten 7200 RPM 1 TB disks. The platform is distributed across three racks. All the switches in the network topology can be considered "non-blocking": all machines can communicate at 1Gbps with each other.

We operate our platform using OpenStack [22], which can automatically provision virtual analytic clusters composed by VMs connected directly to each other (i.e., no traffic encapsulation and no virtual routing).

Our platform provides volumes (similar to Amazon EBS [23]) and ephemeral disks Storage layers. Volumes are provisioned through the Openstack's Cinder module [24] on top of Ceph [25] that is a distributed file system featuring high performance, reliability and scalability. Ceph's blocks are distributed over 8 disks spread across 5 physical machines. Similarly to a traditional RAID 0 approach, when performing read and write operations, Ceph divides the data in smaller chunks (8 MB) and store them across storage servers, called Object Storage Daemons (OSDs). Instead, ephemeral disks are connected to a portion of a single physical disk that reside on the same host that runs the virtual machine using it.

Our private cloud uses the OpenStack Sahara project to automatically provision Compute layers: in this work, we use Apache Spark [7]. Spark can read from several Data layers; the two widespread solutions that we take into consideration are HDFS and Swift (which is similar to Amazon S3 [8]). HDFS is a Java-based distributed file system providing scalable and reliable data storage, designed to store a small number of large files. Swift is a highly available, distributed, eventually consistent object store designed as a generic service to reliably store very large numbers heterogeneous files. In our platform, Swift is deployed on a single physical machine using the Swift-All-in-One (SAIO [6]) configuration, on the basis of the capacity planning suggested in [26]; no other processes share Swift's hardware. More generally, Swift can be deployed on several machines, to increase, e.g., capacity and reliability, at the cost of increased network traffic. It is worth noting that we disabled the Swift authentication mechanism in order to avoid additional overhead in the communication process, and to focus only on the data path.

In this work, we gloss over the intricacies due to multi-tenancy and interference: hence, we statically allocate a portion of the platform to run our experiments. The Compute layer is virtualized on 5 VMs and uses Spark 1.5.2 as the computing framework: 4 workers, spread across 4 different hosts, and 1 master. The Data layer is also virtualized in 5 VMs spread across 5 different hosts. All the VMs are equipped with *(i)* 4 cores, *(ii)* 32 GB of RAM and *(iii)* 80 GB of disk.

To gain statistical confidence in our results, all the experiments we report in this article were repeated five times.

### B. Deployment scenarios

We define a deployment scenario as a configuration of Compute, Data and Storage layers and study 4 scenarios that we think representative of common configurations:

*1) GC:* Guest Collocation. The Compute and Data layers are hosted on the same VM, the Storage layer is an ephemeral, local disk. This is a popular configuration in public clouds: when a GC cluster is decommissioned, all data is lost.

*2) GC-V:* Guest Collocation with Volumes. Same configuration as for GC, but the Storage layer uses volumes provisioned using the Ceph distributed file system. This is also a popular configuration in public clouds: it enables elasticity at the Compute layer, without sacrificing data durability at the Data and Storage layers.

*3) SWI:* Swift. The Compute and Data layers run on distinct hosts. The Data layer is the Swift object store, which ensures data durability. Similarly to the GC-V configuration, this scenario enables Compute layer elasticity, and it is a popular configuration for its simple REST-based interface to interact with data.

*4) NC:* No Collocation. The Compute and Data layers run on different hosts; the Data layer uses HDFS mounted on a Storage layer that uses ephemeral, local disk. This is a scenario enabling data durability: the Compute layer can be decommissioned, while the Data and Storage layer keep running.

As we discuss in Section IV-C, each scenario presents a different degree of data locality. In addition, we note a first ex-ample of impedance mismatch, which we further elaborate in the remainder of this work. Indeed, both the Data and Storage layers might implement their own data replication mechanism. This is evident in the GC-V scenario: when HDFS is mounted on volumes, HDFS and Ceph replication mechanisms are redundant. To better understand the performance implications of the GC-V scenario, we thus distinguish two cases: GC-Vb, with both HDFS and Ceph replication and GC-Vs, with only Ceph replication, which correspond to the degrees of freedom exposed to users for configuring their services. Note that if HDFS replication is disabled, Spark (the Compute layer) has no other means to retrieve the data if a datanode crashes, resulting in a failure of the application; also Spark's scheduler has less flexibility in scheduling tasks, because data blocks are present in only one datanode. In contrast, by enabling HDFS replication, the application will write extra data.

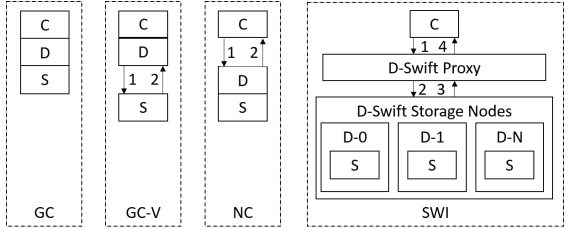

Fig. 2. *Compute-to-Data path* for different scenarios during read operations.

### C. Compute-to-Data path

We now introduce an intuitive notion of distance between where computation happens and data reside. As illustrative examples, consider the following cases: compute and data reside on the same VM, on different VMs running in the same physical host, on different VMs on different physical hosts in the same rack, and so on. It is intuitive to treat these cases as increasing in terms of distance, which is thus loosely coupled with the amount of network links data need to traverse for being processed. Also, recall that read and write operations issued by the Compute layer, work on a given split of the input or output data, which is organized as a sequence of records. We use the following intuitive and rough definition of distance:

*Definition:* The *Compute-to-Data path* is the number of logical links a successful (read or write) operation must cross, for a given data record.

In this Section we use the *Compute-to-Data path* as a proxy to reason about performance ranking, that takes into account the logical distance between the three layers composing an analytic service and the additional cost of the replication system(s). Indeed, we can expect a performance degradation each time an operation traverses a network link: the intuitive ranking holds even if we do not explicitly model network latency or topology.

Figure 2 shows graphically how we derive the *Compute-to-Data path* for each scenario during a **read request**. Note that it is important to be careful and take into account the architecture details of each layer: for example, Swift has a single point of access called the Swift-Proxy, which mediates between the Compute layer and Swift's storage nodes. To calculate the *Compute-to-Data path* we count all the logical links, between the Compute layer and the physical data, that each individual record request has to traverse. Considering the SWI scenario, we have one link from Compute layer to Swift-proxy, then one more between Swift-proxy and Swift' storage nodes, finally the record request's ACKs traverse the same links to reach the Compute layer, for a total of 4 links. Similar consideration can be made for the remaining scenarios: the GC traverses 0 link while the GC-V and the NC 2 links.

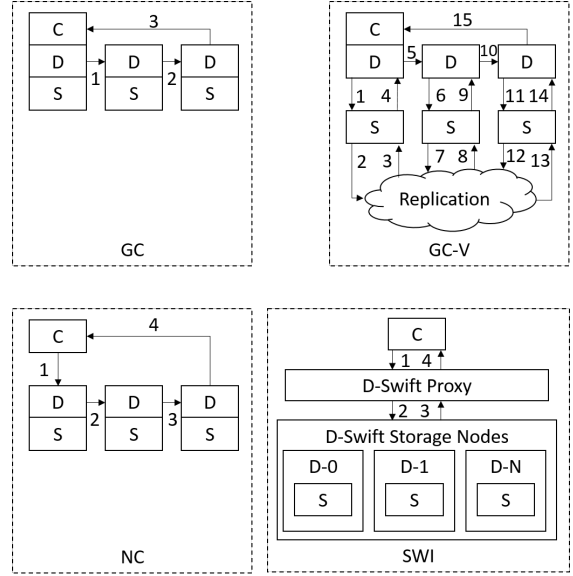

Fig. 3. *Compute-to-Data path* for different scenarios during write operations.

For **write requests**, data replication has to be taken into account. HDFS, Swift and Ceph have different replication systems: HDFS uses *chain-replication*, whereas Swift and Ceph use *asynchronous replication*, with different quorums. Assuming a replication factor of 3, the Swift-proxy requires 2/3 of storage nodes to acknowledge a write operation, whereas Ceph requires all OSDs to acknowledge success. Figure 3 illustrates how to derive the *Compute-to-Data path* for each scenario during a write request. Taking GC-Vb as example: a single datanode record write operation traverse 4 links, since we have HDFS replication active with a factor of 3, we will have 12 links; we also have to count the links between datanodes and the final ACK, for a total of 15 links.

The *Compute-to-Data path* for read and write operations, and for different scenarios is summarized in Table I. In the Table, we organize and rank scenarios based on their distance: intuitively, we expect application performance to follow the same ranking we produce using the *Compute-to-Data path*. Our measurement results indicate that the *Compute-to-Data path* is a good proxy to rank scenarios based on their expected relative performance, albeit intuition is not sufficient alone to explain what we support with data.

TABLE I
EXPECTED SCENARIOS' PERFORMANCE RANKING.

(a) Read

| Rank | Scenario | #links |
|---|---|---|
| 1 | GC | 0 |
| 2 | GC-V | 2 |
| 2 | NC | 2 |
| 3 | SWI | 4 |

(b) Write

| Rank | Scenario | #links |
|---|---|---|
| 1 | GC | 3 |
| 2 | GC-Vs | 4 |
| 2 | NC | 4 |
| 2 | SWI | 4 |
| 3 | GC-Vb | 15 |

TABLE II
WORKLOADS' DETAILS.

| Workload | #Jobs | #Mapper | #Reducer | Input Size | Output Size |
|---|---|---|---|---|---|
| WordCount | 1 | 158 | 158 | 20 GB | 225.6 MB |
| DFSIO | 1 | 160 | 0 | 0 B | 20 GB |
| TPC-DS | 217 | 16160 | 16821 | 13 GB | 17.9 KB |
| Decision-Tree | 153 | 4549 | 4825 | 3 GB | 8 MB |

*D. Benchmark and Workloads*

To study the performance of analytics applications we use four workloads (described in details in Table II), that are currently used in popular benchmark tools suites and cover different kinds of applications. We use WordCount and DFSIO from Intel Hi-Bench [27], [28] test suites; the first is the "Hello World" application for parallel computing, which is a read-intensive workload, while the second is a write intensive application: they both perform read and write operations on plain-text files. TPC-DS is a transaction processing performance Council's decision-support benchmark test [29], [30], by DataBricks' Spark-Sql-Perf library [31], that executes 5 complex queries[2] from files stored using the Parquet Format [32]. Decision-Tree is a machine learning algorithm taken from Spark's MLlib library [33], [34] that reads CSV files and builds a statistical model of the underlying data distribution; this is the only workload that uses *caching* for the input data: due to its iterative nature, this is the current best-practice to achieve low training times.

*E. Performance metrics*

To investigate the impact that different configurations have on application performance, we use 4 metrics. The first is the *job runtime*: this is the amount of time required by the application to terminate its execution. To delve into the reasons behind each workload's behavior in each scenario, we define extra metrics collected for each analytics application during its execution. These metrics are the percentages of CPU, Network and Disk used by the application itself, computed by standard tools such as *iostat* [35].

To compute the above metrics, we monitor each component of the clusters deployed for a specific scenario: Spark master, Spark worker, HDFS namenode, HDFS datanode and Swift[3]

---

[2]In Databricks' library they are called *simple-queries*.
[3]We monitor the Swift-All-in-One deployment as a single component.

## V. RESULTS

In the following we analyze the performance of each workload and its behavior on each scenario, and summarize our findings, discussing the implications for both end-users and providers of cloud services. Furthermore, we discuss about possible directions to mitigate the performance degradation that some Data and Storage layers incur.

*A. Analytics Application Benchmark*

Application performance is easier to understand when results are grouped by workload type. For this reason, in what follows we first delve into the details of each application we use in our experiments. The template we use in our analysis is as follows: first, we discuss whether the ranking produced by our intuitive *Compute-to-Data path* matches that of real workloads, then provide experimental evidence to explain outliers[4].

*1) WordCount:* In general, we remark that the expected rank produced in Table I is representative of application performance: from Table IIIa we see that the GC and GC-V scenarios have roughly the same performance, whereas the NC and SWI scenarios are slower.

The NC scenario constitutes an interesting out-lier: application runtime is roughly 25% slower compared to higher rank scenarios. This is caused by a low CPU utilization that is a direct effect of an inefficient use of network resources: indeed, all read/write operations are synchronous, thus the CPU is blocked until the operation is completed. Note that, although the NC and GC-V cases have the same *Compute-to-Data path*, their data access mechanism is different. In the NC scenario, when a Compute instance requests a record from a datanode, the record is read over a single disk and network link, which performs poorly overall. Furthermore, aside from being a slow configuration, the NC scenario is also the most expensive one.

Instead, the GC-V scenario achieves very good performance, even if the network cards we use in our platform (1 Gbps interfaces) are not on par with what is currently deployed in public clouds such as AWS (10 Gbps interfaces). This is the result of parallel data transfers, which use network resources more efficiently: fragments of data records are read or written by several disks and transfered over multiple network links. As such, the expected performance degradation caused by a large *Compute-to-Data path* is practically nullified. Additionally, we remark that application performance can reap the benefits of having both Data and Storage layer replication enabled (GC-Vb): indeed, Spark's scheduler has more flexibility in choosing the designated executor for a specific task since there are multiple copies of the same data block.

As expected, the SWI configuration achieves the worst performance: as we discuss later, this is due to both poor architectural choices (Swift was not originally designed to serve parallel processing frameworks) and to problems that arise between Spark and Swift.

---

[4]Not all figures will be shown due to space constraint.

TABLE III
ANALYTICS APPLICATIONS BENCHMARK RESULTS IN ASCENDING ORDER.

(a) WordCount

| Rank | Scenario | Run Time (s) |
|---|---|---|
| 1 | GC-Vb | 121.29 ± 2.20 |
| 1 | GC | 125.23 ± 2.15 |
| 1 | GC-Vs | 125.28 ± 1.89 |
| 2 | NC | 157.85 ± 2.94 |
| 3 | SWI | 279.55 ± 4.07 |

(b) TPC-DS

| Rank | Scenario | Run Time (s) |
|---|---|---|
| 1 | GC-Vb | 454.48 ± 6.89 |
| 1 | GC | 460.21 ± 3.95 |
| 1 | GC-Vs | 469.66 ± 9.31 |
| 2 | NC | 571.01 ± 3.98 |
| 3 | SWI | 2773.96 ± 16.89 |

(c) DFSIO

| Rank | Scenario | Run Time (s) |
|---|---|---|
| 1 | GC | 305.86 ± 14.68 |
| 1 | NC | 308.83 ± 13.38 |
| 1 | GC-Vs | 330.99 ± 13.15 |
| 2 | GC-Vb | 848.48 ± 60.74 |
| 3 | SWI | 1114.56 ± 28.22 |

(d) Decision Tree

| Rank | Scenario | Run Time (s) |
|---|---|---|
| 1 | GC-Vb | 997.50 ± 16.47 |
| 2 | NC | 1067.35 ± 33.48 |
| 2 | GC | 1076.68 ± 39.74 |
| 2 | SWI | 1101.37 ± 21.74 |
| 2 | GC-Vs | 1133.56 ± 37.13 |

*2) TPC-DS:* Table IIIb indicates that there is a good match between the expected ranking obtained using the *Compute-to-Data path* (TPC-DS is a read intensive workload) and application runtimes. Similarly to the WordCount workload, the NC scenario suffers from inefficient use of network resources. Figure 5, which shows the CDF of the task runtime for each workload, supports this claim: tasks are slower in the NC configuration, compared to the GC and GC-V scenarios.

To better understand the causes of the increased tasks runtime, in the SWI scenario, we study resource utilization. Figure 4 shows the resource utilization across different scenarios. In SWI, the network may be considered the source of slowdown (as shown by the median at almost 100%): in fact, in Swift all requests to read or write data records pass through the same physical channel (at the proxy), stressing the network and its chances to become a bottleneck. To tackle this issue, AaaS providers adopt two approaches: faster network links (e.g., 10 Gbps) and deployment of several proxies to balance the traffic. These solutions partially mitigate the problem. Using faster network links may work, but does not scale with the number of concurrent tenants. Using several proxies and a load balancer introduces other problems: *(i)* additional delays due to an extra communication step; *(ii)* a single storage node will have to handle more requests coming from different proxies, thus moving the pressure from the link *Compute-to-Proxy* to the link *Proxy-to-Storage nodes*. Inevitably, additional proxies bring Swift architecture closer to that of HDFS-like system, but at an higher cost in terms of required hardware and consequently, requiring a difficult capacity planning. Additional factors that contribute to performance slowdown include the extra time required by Swift to parse HTTP requests and to dispatch them to the different storage nodes. Finally, from Fig. 4 we can also see how the CPU utilization in the Compute layer of the SWI scenario is lower compared to the other scenarios; similarly to WordCount, this is because the Compute layer has to wait more time to process the data.

Clearly, a narrow measurement campaign that focuses on a single scenario (e.g., the GC configuration) might lead to inaccurate conclusions: even if the workload may be considered CPU-bound – thus suggesting data locality to be irrelevant – different configurations with different levels of data locality have a non-negligible impact on application runtimes.

*3) DFSIO:* Table IIIc shows the measured runtime of DFSIO, which is a write intensive workload, for several

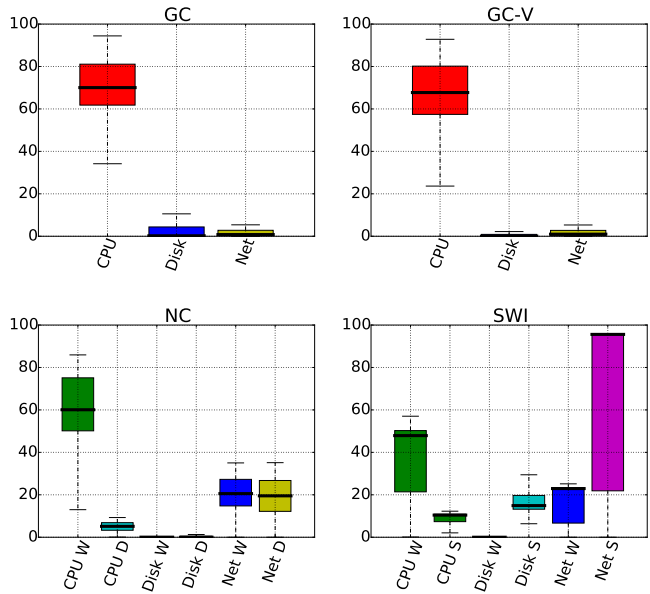

Fig. 4. Resource utilization with the TPC-DS workload in different scenarios. The ticks on the X axis "W", "D" and "S" stand, respectively for worker, datanode and Swift machines. The resource utilization reported is a global average across all instances of each layer. The network utilization between the GC-V and GC scenarios is similar because volumes' network is opaque to the Operating System and, therefore, counted in the disk utilization.

configurations. As expected, the intuitive ranking we derive using the *Compute-to-Data path* only partially matches with experimental results: impedance mismatch between layers is the main culprit for performance degradation.

First, we focus on the SWI scenario: to better understand our experimental results we run a micro benchmark that, using the *python-swift*[5] client, emulates the DFSIO workload: writing the same amount of data as for the DFSIO workload takes only 300 seconds. A detailed log inspection indicates that Swift – because it stores immutable objects with immutable identifiers – does not work well in conjunction with current parallel processing frameworks such as Spark. Indeed, Spark tasks always output temporary files that are renamed once the processing is complete[6]. Since in Swift a **rename** operation is implemented as a **copy** operation, the DFSIO workload in the SWI configuration involves writing much more data than required, hence the poor performance. In particular the data are written 2 extra times, the first from tasks themselves and a second time from the Spark master when the job is completed. From the logs we can see that the job wrote 20 GB in 600 seconds (s) while the application ran in 1114s; in the 600s spent by the job, every task performed a rename operation writing the output one extra time; the difference between the job and the application runtime (514s) corresponds to the time spent by the Spark's master to rename all the files written by the tasks, that is, to write the output a third time. Figure 5 shows the distribution of task runtimes for the DFSIO workload: for the SWI scenario, tasks are much slower than in other scenarios, which corroborates our claims[7].

---
[5]https://github.com/openstack/python-swiftclient
[6]This is reminiscent of the failure tolerance mechanisms in Spark, that assumes tasks can fail at any time.

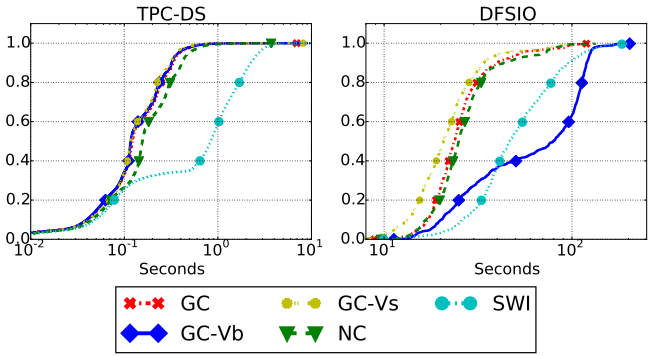

Fig. 5. DFSIO and TPC-DS CDFs for task runtimes in all scenarios.

Next, we focus on the GC-V scenarios. In this case, the impedance mismatch between Data and Storage layers is to blame for poor workload performance. Table IIIc indicates that by disabling HDFS replication mechanism (scenario GC-Vs), performance drastically improves when compared to a naive deployment with both replication mechanisms (HDFS and Ceph) enabled. Although this is not possible to study in a public cloud, we also run experiments in which we disable Ceph replication and only use HDFS replication: in this case, the application runtime falls to roughly 250 seconds, that it is 20% faster to complete with respect to the GC scenario. In summary, write-intensive applications can be significantly affected by impedance mismatch between services: replication and failure tolerance mechanisms implemented in different layers must be tuned to achieve better performance.

*4) Decision-Tree:* The decision tree algorithm taken from Spark's MLlib is at the heart of our machine learning use case: it is an iterative algorithm that builds a statistical model using millions of training data points. Table IIId indicates that application runtimes are essentially independent from system configurations. This is the results of using Spark caching mechanism. When using caching at the application level, the interaction between Compute and Data/Storage layers is lim-ited to reading the input data; the bulk of the computation, and hence the application runtime, happens during the iterations of the decision tree algorithm, in which the Data and Storage layers do not intervene. Finally, the output of the algorithm is a statistical model, which is very small for this workload.

---

[7]We are aware of a work from IBM (https://github.com/SparkTC/stocator) that aims to address the file renaming problem.

Finally, we note that the GC-Vb scenario performs better than others: indeed, *(i)* Ceph reads from multiple disks and *(ii)* Spark's internal scheduler has more flexibility in placing tasks because of the increased data redundancy at the HDFS layer. In general, the most used resource is CPU, while network and disk I/O are barely used.

### B. Summary of the results

Choosing the right composition of analytic services is a difficult problem, involving cost considerations, data durability requirements, and ultimately, expected application performance. Our experimental findings pave the way to informed decisions about AaaS deployments. In the following, we summarize our results and their implications.

**Service composition.** A configuration that aims at achieving data durability in spite of the ephemeral nature of VMs and the services they execute, must be designed with care. For reasons ranging from ease of integration to familiarity with well-established APIs, it is tempting to compose services as done in the NC (no collocation) scenario we study in this work. Our results show that this is a bad choice for a wide range of workloads, in which precious CPU cycles are lost to wait for data to travel over the network.

**Volumes.** Using volumes provisioned on top of a distributed file system like Ceph perform surprisingly well. This is unexpected as, similarly to the NC scenario, the network is heavily involved during application execution. However, our results indicate that even with a modest bisection bandwidth, the Compute layer can make quick progress toward the end of an application, thanks to the efficiency of striping.

However, as cautionary note, our results indicate a potential impedance mismatch between Data and Storage layers, due to the interaction of multiple replication mechanisms. As such, end-users should be aware of the situation, and appropriately configure the Data layer, such that data replication is only performed by the Storage layer, because this is of great benefit to application performance.

Additionally, our results indicate that cloud computing providers could differentiate their volume offering: general purpose volumes would work as usual, whereas analytics volumes should disable data replication. In this case, end-users would be in complete control of replication: our results show that – especially for write intensive workloads – this produces superior performance.

**Swift.** The performance of Swift is disappointing. This is due to another instance of impedance mismatch, between the Compute and Data layer. Swift inability to rename files without actually creating a new copy, causes severe performance penalties, making Swift a sub-optimal solution.

In addition, we note that the Swift architecture was designed for applications that are very different from parallel computing frameworks: our results indicate that the Swift proxy may represent a bottleneck, since it is involved in the data path. Certainly, using a proxy server as coordinator enables cluster managers to easily add control flows to Swift, but this degrades performance. One solutions that is currently adopted by several companies using Swift, at a production level, is to add several proxies and balance the traffic load between them: but because the workload may change unpredictably, a well thought capacity plan is not easy to obtain. As previously underlined, more proxies will make the Swift's architecture somehow similar to HDFS, but at higher costs. Recent work from IBM [36] shows that some control flow and data transformations can be done much closer to the storage nodes. As a consequence, it is tempting to suggest the design of a new Swift proxy that could behave similarly to the HDFS NameNode: such alternative proxy would only act as a metadata storage, and would not be involved in the actual data path.

**Caching.** Finally, our results show that caching plays an important role in determining application performance. On the one hand, caching "breaks" the *Compute-to-Data path* that can be inferred from read/write operations on data records, which makes application performance more difficult to predict. On the other hand, by collapsing the *Compute-to-Data path*, it mitigates the problems of several configurations we studied, which is helpful for end-users because it gives more flexibility in choosing Data and Storage layers.

However, the design of inter-application caching mechanism for parallel processing frameworks is still in its infancy: Tachyon [37] and HDFS2 are good examples of recent approaches to tackle this problem.

## VI. CONCLUSIONS AND FUTURE WORK

We investigated the impact of different Compute, Data and Storage layer configurations on the performance of a data analytic framework. We took an experimental approach, and proposed a measurement campaign, whose objective was to analyze workload performance in light of an intuitive notion of distance between where computation happens and data reside. First, we discussed how to approximately rank different service compositions, in terms of expected performance. Then we performed an extensive measurement campaign on a private cloud computing environment. Results indicated that, in general, our intuitive distance metric is a good proxy to reason about performance ranking. Finally we presented experimental evidence of the impedance mismatch that affect two important storage layers – object and elastic block stores – and deduced mechanism to mitigate negative effects on performance.

As for many measurement studies, it is reasonable to question the generality of the conclusions, when experiments are performed on a single (and very complicated) platform instance. To this end, our research road map includes extending our measurement methodology to public cloud providers, such as AWS, and consider additional workloads. For example, we expect a service like Amazon S3 to achieve superior performance than Swift. Note that our intuitive methodology to approximately rank scenarios nicely extends to large cloud providers, where a detailed view of system internals – that could define a more accurate distance metric – is not available.

ACKNOWLEDGEMENTS

The research leading to these results has received funding from the H2020-644182 project "IOStack".